\documentclass[11pt]{amsart}

\usepackage{amsmath,amssymb,dsfont}

\usepackage{texdraw}

\newfont{\bbd}{msbm10 scaled\magstep1}

\def\id{\hbox{{1}\kern-.25em\hbox{\rm l}}}
\def\one#1{#1^{\raise5pt\hbox{$\scriptstyle\!\!\!\!1$}}\,{}}
\def\two#1{#1^{\raise5pt\hbox{$\scriptstyle\!\!\!\!2$}}\,{}}

\def\comment#1{}

\def\?{(?)\marginpar{|?}}

\def\beq{\begin{equation}}
\def\eeq{\end{equation}}
\def\bea{\begin{eqnarray}}
\def\eea{\end{eqnarray}}
\def\bmat{\left(\begin{array}}
\def\emat{\end{array}\right)}

\newtheorem{theorem}{Theorem}

\makeatletter
\expandafter\let\expandafter
\reset@font\csname reset@font\endcsname
\def\subeqnarray{\arraycolsep1pt
    \def\@eqnnum\stepcounter##1{\stepcounter{subequation}%
        {\reset@font\rm(\theequation\alph{subequation})}}
\jot5mm     \eqnarray}

\makeatother

\newcounter{appendix}
\def\l{\lambda}

\def\mat2#1#2#3#4{{\left(\begin{array}{cc}#1 & #2\\ #3 & #4
      \end{array}\right)}}

\def\mats2#1#2#3#4{{\left(\begin{array}{cc}#1 & #2\vspace{2truemm} \\ #3 & #4
\end{array}\right)}}

\def\tilde{\widetilde}

\begin{document}

\title[The Lattice Schwarzian KdV equation and its Symmetries]
{\bf The Lattice Schwarzian KdV equation and its Symmetries}

\author{DECIO LEVI}

\address{Dipartimento di Ingegneria Elettronica \\
Universit\`a degli Studi Roma Tre and Sezione INFN, Roma Tre \\
Via della Vasca Navale 84, 00146 Roma, Italy\\
E-mail: levi@roma3.infn.it}

\author{MATTEO PETRERA}

\address{Zentrum Mathematik \\
Technische Universit\"at M\"unchen \\
Boltzmannstr. 3,
D-85747 Garching bei M\"unchen,
Deutschland\\
Dipartimento di Fisica E. Amaldi \\
Universit\`a degli Studi Roma Tre and Sezione INFN, Roma Tre \\
Via della Vasca Navale 84, 00146 Roma, Italy\\
E-mail: petrera@ma.tum.de}

\author{CHRISTIAN SCIMITERNA}

\address{Dipartimento di Fisica E. Amaldi \\
Universit\`a degli Studi Roma Tre and Sezione INFN, Roma Tre \\
Via della Vasca Navale 84, 00146 Roma, Italy\\
E-mail: scimiterna@fis.uniroma3.it}

\begin{abstract}
 In this paper we present a set of results  on the symmetries of the lattice Schwarzian
Korteweg-de Vries (lSKdV) equation. We construct the Lie point symmetries and, using  its associated spectral problem,
 an infinite sequence of generalized symmetries and master symmetries. 
We finally show that we can use  master symmetries of the lSKdV equation to construct 
 non-autonomous non-integrable generalized symmetries.  
\end{abstract}

\maketitle

\section{Introduction}
The lattice version of the Schwarzian Korteweg-de Vries (lSKdV) equation
\bea \label{pkdv}
w_t=w_{xxx}-\frac{3 \,w_{xx}^2}{2 \, w_x},
\eea
is given by the nonlinear partial difference equation \cite{frank,nqc}:
\bea
&& \mathbb{Q} \doteq \alpha_1 \, (x_{n,m}-x_{n,m+1}) \, (x_{n+1,m}-x_{n+1,m+1})-
\label{q1} \\
&& \qquad \,   - \, \alpha_2 \, (x_{n,m}-x_{n+1,m})\, (x_{n,m+1}-x_{n+1,m+1})=0.
\nonumber
\eea
Eq. (\ref{q1})  involves just four lattice
points which lay on two
orthogonal infinite lattices and are situated at the vertices of an elementary square.
It is a lattice equation on {\it quad-graphs} belonging
to the recent classification presented in \cite{abs}, where the $3D$ {\it consistency} is used as a tool to establish
its integrability. It is a subcase of  the first element of the {\it Q--list}, 
namely the equation $Q_1$, with $\delta = 0$.

As far as we know the lSKdV equation has been introduced for the first time by Nijhoff, Quispel and Capel  \cite{nqc} in 1983.  A review of results about the lSKdV equation can be found in \cite{frank1,frank}.

Since we have two discrete independent variables, i.e. $n$ and $m$,
we can perform the continuous limit in two steps. Each step is   achieved by shrinking the
corresponding lattice step to zero and sending to infinity the number of points of the lattice.

In the first step, setting $\alpha_1\doteq q^{2}$, $\alpha_2\doteq p^{2}$,  we
define $x_{n,m} \doteq {\tilde x}_{k}(\tau)$, where
$k \doteq n+m$ and $\tau \doteq \delta \,  m$, being  $\delta \doteq p-q$.
Considering the limits 
$m \rightarrow \infty, \, \delta \rightarrow 0$, we get the
differential-difference equation
\beq
\frac{d \,{\tilde x}_{k}(\tau)}{ d \, \tau} =\frac{2\, ({\tilde x}_{k+1}-{\tilde x}_{k})\,
({\tilde x}_{k-1}-{\tilde x}_{k})}{p\, ({\tilde x}_{k-1}-{\tilde x}_{k+1})}.\label{2}
\eeq

The second continuous limit of eq. (\ref{q1}) is performed by taking
in eq. (\ref{2}) ${\tilde x}_{k}(\tau)\doteq w(x,t)$, with 
$x\doteq 2 \, (k + \tau / p )/p$ and $ 
t\doteq 2 \, ( k/3+ \tau /p ) /p^3.$
If we carry out the limit 
$
p \rightarrow \infty, \, k \rightarrow \infty, \, \tau \rightarrow \infty,
$ in such a way that $x$ and $t$ remain finite, then  
eq. (\ref{2}) is transformed into the continuous Schwarzian KdV equation  (\ref{pkdv}).

Integrable equations possess an infinite set of symmetries. Few of them are {\it point symmetries}, i.e. 
symmetries whose infinitesimal generators depend just on the independent and dependent variables, while an
infinite denumerable number of them 
are {\it generalized symmetries}. These latter ones depend also on the derivatives of the dependent variable, 
for the continuous independent variables, and on a few lattice points, if the independent variables are discrete. 
The presence of this infinite Lie algebra of symmetries is one of the most important features of the
integrability of a given nonlinear equation and it has been used with profit in the past to provide
integrability tests for several partial differential equations in 
$\mathbb{R}^2$ and differential-difference
equations\cite{yamilov1,yamilov2}. 

To be able to introduce an integrability test based on symmetries, one needs to understand the structure of the 
infinite dimensional symmetry algebra of the integrable equations. 
In the case of completely discrete equations the situation is not as clear as 
for the differential-difference or the partial-differential case.
Results in this direction have been obtained some time ago
for the discrete-time Toda lattice \cite{martina} and recently for the lattice potential Korteweg-de Vries equation \cite{lp}.  
However, the Toda Lattice and the lattice potential KdV equation are just examples and more examples are needed to get a sufficiently general 
idea of the possible stuctures which may appear. 

The present paper is part of this research and is devoted to the study of the lSKdV equation exactly from this point of view.
In section \ref{S1} we present  some old and new  results on the integrability of the lSKdV equation. 
Section \ref{S2} is devoted to the construction of its Lie point symmetries
while in section \ref{S3} we consider its generalized  symmetries. Finally,
section \ref{S4} contains some concluding remarks.

\section{The integrability of the lSKdV equation} \label{S1}

Eq. (\ref{q1}) has been obtained firstly by the direct linearization method
\cite{nqc}.  In \cite{frank} one can find its associated spectral problem, which, as this equation 
 is part of the 
Adler-Bobenko-Suris classification, can be obtained using a well-defined procedure
\cite{bs1,bs2,n}. 

Its Lax pair is
given by the following overdetermined system of matrix 
equations for the vector $\Psi^\l_{n,m} \doteq
(\psi^1_{n,m}(\l),\psi^2_{n,m}(\l))^T$,
$\lambda \in \mathbb{R}$,
\begin{subequations} \label{q1lax}
\begin{align}
&\Psi^\l_{n+1,m}=L_{n,m}^\l \, \Psi^\l_{n,m}, \label{q1lax1} \\
&\Psi^\l_{n,m+1}=M_{n,m}^\l \, \Psi^\l_{n,m}, \label{q1lax2}
\end{align}
\end{subequations}
where
$$
L_{n,m}^\l \doteq \left(\begin{array}{cc}
1 & x_{n,m}- x_{n+1,m}  \\
\l \, \alpha_1 \, (x_{n,m}- x_{n+1,m})^{-1} & 1
\end{array}\right),
$$
and
$$
M_{n,m}^\l \doteq \left(\begin{array}{cc}
1 & x_{n,m}- x_{n,m+1}  \\
\l \, \alpha_2 \, (x_{n,m}- x_{n,m+1})^{-1}  & 1
\end{array}\right).
$$
The consistency of eqs. (\ref{q1lax}) implies the discrete
Lax equation
\beq \nonumber
L_{n,m+1}^\l \, M_{n,m}^\l=M_{n+1,m}^\l \,L_{n,m}^\l.
\eeq
We can rewrite eqs. (\ref{q1lax}) in scalar form in terms
of $\psi_{n,m} \doteq \psi^1_{n,m}(\l)$:
\begin{subequations} \label{Lax12}
\begin{align}
&(x_{n,m}-x_{n+1,m})\, \psi_{n+2,m}+
(x_{n+2,m}-x_{n,m})\, \psi_{n+1,m}+ \nonumber \\
& \qquad +\left(1-\l \, \alpha_1 \right)
(x_{n+1,m}-x_{n+2,m})\, \psi_{n,m}=0, \label{lax3}\\
&(x_{n,m}-x_{n,m+1})\, \psi_{n,m+2}+
(x_{n,m+2}-x_{n,m})\, \psi_{n,m+1}+ \nonumber \\
& \qquad +\left(1-\l \, \alpha_2\right)
(x_{n,m+1}-x_{n,m+2})\, \psi_{n,m}=0.\label{lax4}
\end{align}
\end{subequations}

It is worthwhile to observe here that  eq. (\ref{q1}) and the Lax equations
 (\ref{Lax12}) are
invariant under the  discrete symmetry obtained by interchanging at the same time $n$ with $m$ and $\alpha_1$ with $\alpha_2$. 

To get meaningful Lax equations the field $x_{n,m}$ cannot go asymptotically to a constant 
but must be written as $x_{n,m} \doteq u_{n,m} + \beta_0 \, m + \alpha_0 \, n$,
where $\alpha_0$ and $\beta_0$ are constants related to $\alpha_1$ and
$\alpha_2$ by  the 
condition
$
\alpha_1 \, \beta_0^2 = \alpha_2 \, \alpha_0^2. 
$
Under this transformation of the dependent variable, 
the lSKdV equation and its Lax pair can be respectively rewritten as
\bea
&& \alpha_1 \, (u_{n,m} - u_{n,m+1} - \beta_0) \, ( u_{n+1,m} - u_{n+1,m+1} -
\beta_0) = \nonumber \\ 
&&  = \alpha_2 \, (u_{n,m} - u_{n+1,m} - \alpha_0) \, ( u_{n,m+1} - u_{n+1,m+1} -
\alpha_0), \label{q1a}
\eea
and
\begin{subequations} \label{Lax1}
\begin{align}
& (1 + v_{n,m}^{(n)}) \, \psi_{n+2,m} - ( 2 + v_{n,m}^{(n)}) \, \psi_{n+1,m} + (1-
\lambda \, \alpha_1) \, \psi_{n,m} = 0, \label{lax3a} \\ 
& (1 + v_{n,m}^{(m)}) \, \psi_{n,m+2} - ( 2 + v_{n,m}^{(m)}) \, \psi_{n,m+1} + (1-
\lambda \, \alpha_2) \, \psi_{n,m} = 0, \label{lax4a}
\end{align}
\end{subequations}
where
\bea \nonumber
v_{n,m}^{(n)} \doteq  \frac{u_{n+2,m} - 2 \, u_{n+1,m} + u_{n,m}}{u_{n+1,m} -
  u_{n,m} + \alpha_0}, \quad v_{n,m}^{(m)} \doteq  \frac{u_{n,m+2} - 2 \, u_{n,m+1} + u_{n,m}}{u_{n,m+1} - u_{n,m} + \beta_0},
\eea
where $u_{n,m} \rightarrow c$, with $c \in \mathbb{R}$, as $n$ and $m$ go to infinity.

One can construct the class of differential-difference equations associated
 with
eq.  (\ref{lax3a}) by requiring
 the existence of a set of operators $M_n$ such that \cite{br} 
$$
L_n \, \psi_n = \mu \, \psi_n, \qquad 
\frac{d \psi_n}{dt} = - M_n \, \psi_n, \qquad \frac{d L_n}{dt} = [ \, L_n,
 M_n\, ].
$$
with $L_{n} \doteq  (1+ v_{n,m}^{(n)}(t))\, E^2 - (2+v_{n,m}^{(n)}(t))\, E$. Here
 $E$ is the (positive) shift operator in the variable $n$. If $d \mu /dt = 0$
 the class of differential-difference equations one so obtains will be called {\it
 isospectral}, while if $d \mu/dt \neq 0$ it will be called {\it
 non-isospectral}.

As the field $v_{n,m}^{(n)}(t)$ appears multiplying both $E$ and $E^2$ the
expression of the recursive operator turns out to be extremely complicate,
containing triple sums and products of the dependent fields. So we look for
transformations of the spectral problem (\ref{lax3a}) which reduce it to a simpler form in which the potential will appear just once. There are two different discrete spectral problems involving three lattice points, the discrete Schr\"odinger spectral problem introduced by  Case \cite{case},
\bea \label{laxs1}
\phi_{n-1} + a_{n} \, \phi_{n+1} + b_{n} \, \phi_n = \lambda \, \phi_n,
\eea
which is associated  with the Toda and Volterra differential-difference equations \cite{hlrw} and the asymmetric discrete Schr\"odinger spectral problem introduced by Shabat and Boiti et al. \cite{boiti2,s} ,
\bea \label{laxs2}
\phi_{n+2} = \frac{2\, p}{s_{n}} \, \phi_{n+1} + \lambda \, \phi_n.
\eea
The latter one has been used to solve the so called discrete KdV equation
\cite{narita}.
In eqs. (\ref{laxs1}--\ref{laxs2}) the functions $a_n,b_n,s_n$ may depend
parametrically on a {\it continuous variable} $t$ but also on a {\it discrete variable} $m$.
As all three spectral problems (\ref{lax3a}--\ref{laxs1}--\ref{laxs2}) involve
just three points on the lattice we can relate them by
a gauge transformation $\psi_n \doteq  f_n(\mu) \, g_n(\{u_{n,m} \}) \,
\phi_n$,
where $\{u_{n,m} \} \doteq (u_{n,m},u_{n\pm 1,m},u_{n,m\pm 1},... )$.
These transformations give rise to a Miura transformation  between the
involved fields. For instance,
when we transform the spectral problem (\ref{lax3a}) into the discrete Schr\"odinger spectral problem (\ref{laxs1}) we get
\begin{subequations} \label{m1}
\begin{align} \label{m1a}
& b_n \doteq b_{n,m} = 0, \\ \label{m1b}
& a_n \doteq a_{n,m}= \frac{4 \, (u_{n+1,m} - u_{n,m} +
  \alpha_0)^2}{(u_{n+2,m} - u_{n,m} + 2 \, \alpha_0)\, (u_{n+1,m} - u_{n-1,m}
  + 2 \, \alpha_0)}.
\end{align}
\end{subequations}
If  we transform the spectral problem  given in eq. (\ref{lax3a}) into the asymmetric discrete
Schr\"odinger spectral problem (\ref{laxs2}) the relation between the fields
$s_n \doteq s_{n,m}$ and $u_{n,m}$ is  more involved as it is expressed in terms of infinite products.

We will use in section \ref{S3} the transformation (\ref{m1b}) and the
equivalent one between (\ref{lax4a}) and (\ref{laxs1}) which will define a
field $\tilde a_{m}$ given by eq. (\ref{m1b}) with $n$ and $m$ and $\alpha_0$
and $\beta_0$ interchanged, to build the generalized symmetries of eq. (\ref{q1a}) from the nonlinear differential-difference equations associated with the spectral problem (\ref{laxs1}) \cite{hlrw} with $b_n$ given by eq. (\ref{m1a}).

\section{ Point Symmetries of the lSKdV equation} \label{S2}
Here we look for the Lie point symmetries of eq. (\ref{q1}),
with $\alpha_1 \neq \alpha_2$,
using the technique introduced in \cite{lw06}. The Lie point symmetries we obtain in this way turn out to be the same as those for (\ref{q1a}).

The Lie symmetries of the lSKdV equation (\ref{q1}) 
are given by those continuous
transformations which leave the equation invariant. From the infinitesimal point of view
they are obtained by requiring the infinitesimal invariant condition 
\beq \label{cca2}
\left. {\rm pr} \, \widehat X_{n,m} \, \mathbb{Q}  \, \right|_{\mathbb{Q} =0} =0,
\eeq
where, as we keep the form of the lattice invariant,  
\beq \label{ccb2}
 \widehat X_{n,m} = \Phi_{n,m} ( x_{n,m}) \partial_{x_{n,m}}.
 \eeq
By $ {\rm pr} \, \widehat X_{n,m}$ we mean the prolongation of the infinitesimal generator 
$\widehat X_{n,m}$  to the other three  points appearing in $\mathbb{Q}=0$,
 i.e. $x_{n+1,m}$, $x_{n,m+1}$ and $x_{n+1,m+1}$.

Solving the equation $\mathbb{Q}=0$ w.r.t. $x_{n+1,m+1}$ and substituting it
in eq. (\ref{cca2}) we get a functional equation for $\Phi_{n,m}( x_{n,m})$.
Looking at its solutions in the form
$\Phi_{n,m}(x_{n,m})=\sum_{k=0}^{\gamma}\Phi_{n,m}^{(k)}\,x_{n,m}^{k}$,
$\gamma \in \mathbb{N}$,
we see that  in order to balance the
leading order in $x_{n,m}$, if $\alpha_1 \neq \alpha_2$, $\gamma$ cannot be greater than $2$, and thus must belong to the interval $[\,0,2\,]$. Equating now to zero the coefficients of the powers of $x_{n,m}$,
$x_{n+1,m}$ and $x_{n,m+1}$
we get an overdetermined  system of determining equations. Solving the resulting difference
equations we find that the functions $\Phi^{(i)}_{n,m}$'s, $i=0,1,2$, 
must be constants. Hence
the infinitesimal generators of the algebra of Lie point symmetries are given
by
\beq \nonumber
 \widehat X^{(0)}_{n,m}=\partial_{x_{n,m}},
\qquad \widehat  X^{(1)}_{n,m}=x_{n,m} \, \partial_{x_{n,m}},
\qquad   \widehat X^{(2)}_{n,m}=x_{n,m}^{2}\, \partial_{x_{n,m}}.
 \eeq
The generators $\widehat X^{(i)}_{n,m}$, $i=0,1,2$, span the Lie algebra
$\mathfrak{sl}(2)$:
$$
\left[\,  \widehat X^{(0)}_{n,m},  \widehat X^{(1)}_{n,m}\, \right]= \widehat
X^{(0)}_{n,m},
\qquad 
\left[\,  \widehat X^{(1)}_{n,m},  \widehat X^{(2)}_{n,m}\, \right]= \widehat X^{(2)}_{n,m},
$$
$$
\left[\,  \widehat X^{(0)}_{n,m},  \widehat X^{(2)}_{n,m}\, \right]=2 \,
\widehat X^{(1)}_{n,m}.
$$

We can write down the group transformation by integrating the differential-difference equation
 \bea \label{ddq1}
 \frac{d \, \tilde x_{n,m}(\epsilon)}{d \epsilon} = \Phi_{n,m}(\tilde x_{n,m}(\epsilon)),
 \eea
 with the initial condition $\tilde x_{n,m} (\epsilon = 0) = x_{n,m}$. 
 We get the M\"obius transformation \cite{bs1,bs2,frank}
$$ 
 \tilde
 x_{n,m}(\epsilon_{0},\epsilon_{1},\epsilon_{2})=\frac{(\epsilon_{0}+x_{n,m})\,
 e^{\epsilon_{1}}}{1-\epsilon_{2}\, (\epsilon_{0}+x_{n,m})\,
 e^{\epsilon_{1}}}, 
$$
where the $\epsilon_{i}$'s are  the parameters associated with the infinitesimal
generators
$\widehat X^{(i)}$, $i=0,1,2$.  

We finally notice that, in the case when $\alpha_1=\alpha_2$ eq. (\ref{q1})
simplifies
to the product of two linear discrete wave equations:
$$
(x_{n,m}- x_{n+1,m+1})\,(x_{n+1,m}- x_{n,m+1})=0,
$$
which is trivially solved by taking $x_{n,m}= f_{n \pm m}$ and the Lie point symmetries belong to an infinite dimensional Lie algebra.

\section{Generalized Symmetries of the lSKdV equation} \label{S3}
A generalized symmetry is obtained when the function $\Phi_{n,m}$ appearing in eq. (\ref{ccb2}) 
depends on $\{x_{n,m}\}$ and not only on $x_{n,m}$. A way to obtain it, is 
to look at those  differential-difference equations (\ref{ddq1})
associated with the spectral problem (\ref{laxs1}) which are compatible with
eq.  (\ref{q1}). From eqs. (\ref{m1}) we see that the lSKdV equation can be
associated with the discrete Schr\"odinger spectral problem when $b_{n,m}=0$,
i.e.  when the associated hierarchy of differential-difference
equations is given by the Volterra hierarchy \cite{hlrw}. So, applying the
Miura transformation (\ref{m1}) to the differential-difference equations of
the Volterra hierarchy we can obtain the symmetries of the lSKdV equation. The Miura transformation (\ref{m1})  preserves the integrability of the Volterra hierarchy if $u_{n,m} \rightarrow c$, with $c \in \mathbb{R}$, as $n$ and $m$ go to infinity. 

The procedure to get the generalized symmetries for the lSKdV  is better shown on a specific example, the case of the Volterra equation, an isospectral deformation of the spectral problem (\ref{laxs1}):
\bea \label{vol}
\frac{ d a_{n,m}(\epsilon_0)}{d \, \epsilon_0} = a_{n,m} \, ( a_{n+1,m} - a_{n-1,m}).
\eea
Let us substitute the Miura transformation, given by eq. (\ref{m1b}), into eq. (\ref{vol}) and let us assume that
\bea \nonumber
\frac{ d u_{n,m}(\epsilon_0)}{d  \epsilon_0} = F_{n,m}(u_{n-1,m}, u_{n,m}, u_{n+1,m}).
\eea
Eq. (\ref{vol}) is thus a functional equation for $F_{n,m}$
which can be 
solved as we did in the previous section by comparing powers at infinity or by transforming it 
into an overdetermined system of linear partial differential equations \cite{ac1,ac2}. In this way we get, up to a point transformation,
\bea \label{vol2}
F_{n,m}=\frac{4 \, (u_{n,m}-u_{n-1,m}+\alpha_0)\,
  (u_{n,m}-u_{n+1,m}-\alpha_0)}{u_{n+1,m}-u_{n-1,m}+2 \, \alpha_0}.
\eea
Eq. (\ref{vol2}) is nothing else but eq. (\ref{2}). One can verify that
eq. (\ref{vol2}) is a generalized symmetry of the lSKdV equation (\ref{q1a}) by proving that
$\Phi_{n,m}=F_{n,m}(u_{n-1,m}, u_{n,m}, u_{n+1,m})$  satisfies  eq. (\ref{cca2}). 

If we start from a higher equation of the isospectral Volterra hierarchy
\bea 
\frac{d  a_{n,m} (\epsilon_1) }{d  \epsilon_1} &=& 
a_{n,m} \, [ \, a_{n-1,m} \, (a_{n-2,m} + a_{n-1,m} +a_{n,m} - 4) - \nonumber \\
\nonumber 
&& - \, a_{n+1,m} \, (a_{n+2,m} + a_{n+1,m} + a_{n,m}-4)\, ],
\eea
we get a second generalized symmetry of the lSKdV equation requiring 
$F_{n,m}=F_{n,m}(u_{n-2,m}, u_{n-1,m}, u_{n,m}, u_{n+1,m}, u_{n+2,m})$. It reads
\bea 
F_{n,m} &=& \frac{
  (u_{n,m}-u_{n-1,m}+\alpha_0)\, (u_{n,m}-u_{n+1,m}-\alpha_0)}
{(u_{n+1,m}-u_{n-1,m}+2 \, \alpha_0)^2}  \times \nonumber \\
&& \times \,\left[ 
\frac{ (u_{n+2,m}-u_{n+1,m}+\alpha_0)\, (u_{n-1,m}-u_{n,m}-\alpha_0)}
{u_{n+2,m}-u_{n,m}+2 \, \alpha_0} - \right. \label{2vol1} \\ 
 && \left. \qquad - \frac{(u_{n-1,m}-u_{n-2,m}+\alpha_0)\,
     (u_{n,m}-u_{n+1,m}-\alpha_0)}{u_{n,m}-u_{n-2,m}+2 \, \alpha_0} \right]. \nonumber
\eea
This procedure could be clearly carried out for any equation of the Volterra hierarchy \cite{hlrw} giving a hierarchy of symmetries for the lSKdV equation.

If we consider the non-isospectral hierarchy the only  local equation is \cite{hlrw}
\beq
\nonumber
\frac{d  a_{n,m}(\sigma)}{d  \sigma} = a_{n,m} \, [ \, a_{n,m} - (n-1) \, a_{n-1,m} + (n+2) \,
a_{n+1,m} -4\, ]
\eeq
and it provides, up to a Lie point symmetry, two local equations:
\bea \label{nvol1}
&&\frac{d  u_{n,m}(\sigma_0)}{d  \sigma_0} = u_{n,m} +\alpha_0 \, n, \\
\label{nvol2}
&&\frac{d  u_{n,m}(\sigma_1)}{d  \sigma_1} =\frac{4 \, n \,
  (u_{n,m}-u_{n-1,m}+\alpha_0)\,
  (u_{n,m}-u_{n+1,m}-\alpha_0)}{u_{n+1,m}-u_{n-1,m}+2 \, \alpha_0}.
\eea
Eq. (\ref{nvol1}) is obtained as the multiplicative factor of  an integration
  constant. One can  easily  show that eq. (\ref{nvol1}) is not a symmetry of the lSKdV
  equation but it commutes with all  its known symmetries. Eq. (\ref{nvol2}) is a
  master symmetry \cite{fm}: it does not commute with the lSKdV equation but
  commuting it with eq. (\ref{vol2}) one gets eq. (\ref{2vol1}) and commuting it with
  eq. (\ref{2vol1}) one gets a higher order symmetry. So through it one can
  construct a hierarchy of generalized symmetries of the lSKdV equation. 
  
In the construction of generalized symmetries for the
differential-difference Volterra equation \cite{hlrw} one was able to
construct a symmetry from  the master symmetry (\ref{nvol2}) by combining it
with the second  isospectral symmetry (\ref{2vol1}) multiplied by $t$.  This
seems not to be the case for difference-difference equations. As was showed in
\cite{lp} for the case of the lattice potential KdV equation, there is no  combination of eq. (\ref{nvol2}) with isospectral symmetries which gives us a symmetry of eq. (\ref{q1a}). 

As the lSKdV equation admits a discrete symmetry corresponding to an exchange of $n$ with $m$ and $\alpha_1$ with $\alpha_2$, one can construct another class of generalized and master symmetries by considering the equations obtained from the spectral problem (\ref{laxs1})  in the $m$ lattice variable depending on the  potential  $\tilde a_{m}$. In this way we get:
\bea \nonumber
\frac{d  u_{n,m}(\tilde \epsilon_0)}{d  \tilde \epsilon_0} &=&
\frac{4  \, (u_{n,m}-u_{n,m-1}+\beta_0)\,
  (u_{n,m}-u_{n,m+1}-\beta_0)}{u_{n,m+1}-u_{n,m-1}
+2 \, \beta_0},\\ \nonumber
\frac{d  u_{n,m}(\tilde \epsilon_1)}{d  \tilde \epsilon_1} &=&
\frac{
  (u_{n,m}-u_{n,m-1}+\beta_0)\, (u_{n,m}-u_{n,m+1}-\beta_0)}
{(u_{n,m+1}-u_{n,m-1}+2 \, \beta_0)^2}  \times \nonumber \\
&& \times \,\left[ 
\frac{ (u_{n,m+2}-u_{n,m+1}+\beta_0)\, (u_{n,m-1}-u_{n,m}-\beta_0)}
{u_{n,m+2}-u_{n,m}+2 \, \beta_0} - \right. \nonumber \\ \nonumber
 && \left. \qquad - \frac{(u_{n,m-1}-u_{n,m-2}+\beta_0)\,
     (u_{n,m}-u_{n,m+1}-\alpha_0)}{u_{n,m}-u_{n,m-2}+2 \, \beta_0} \right].
\eea
and
\bea \nonumber 
\frac{d  u_{n,m}(\tilde \sigma_0)}{d  \tilde \sigma_0} &=&  u_{n,m} +\beta_0 \,
m, \\
\nonumber
\frac{d  u_{n,m}(\tilde \sigma_1)}{d  \tilde \sigma_1} &=& \frac{4 \, m \,
  (u_{n,m}-u_{n,m-1}+\beta_0)\,
  (u_{n,m}-u_{n,m+1}-\beta_0)}{u_{n,m+1}-u_{n,m-1}+2 \, \beta_0}.
\eea

A different class of symmetries can be obtained
 applying the following theorem, introduced in \cite{lp}, 
which provides a constructive tool to obtain generalized symmetries  for the lSKdV equation (\ref{q1}).

\begin{theorem}
Let $\mathbb{Q}(u_{n,m},u_{n\pm 1,m},u_{n,m\pm 1},...;\alpha_1,\alpha_2) =0$ be an integrable partial difference equation
invariant under the discrete symmetry $n \leftrightarrow m$, $\alpha_1 \leftrightarrow \alpha_2$.
Let $\widehat Z_n$ be the differential operator
$$
\widehat Z_n \doteq Z_{n} (u_{n,m},u_{n\pm 1,m},u_{n,m\pm 1},...;\alpha_1,\alpha_2) \,  \partial_{u_{n,m}},
$$
such that 
\beq
\left. {\rm pr}\, \widehat Z_n \, \mathbb{Q}\,  \right|_{\mathbb{Q} =0} = 
a \, g_{n,m}(u_{n,m},u_{n\pm 1,m},u_{n,m\pm 1},...;\alpha_1,\alpha_2),
\nonumber
\eeq
where $g_{n,m}(u_{n,m},u_{n\pm 1,m},u_{n,m\pm 1},...;\alpha_1,\alpha_2)$ is invariant under 
the discrete symmetry $n \leftrightarrow m$, $\alpha_1 \leftrightarrow \alpha_2$
and $a$ is an arbitrary constant. Then
\beq
\left. \left( \frac{1}{a}\,{\rm pr}\, \widehat Z_n - 
\frac{1}{b}\, {\rm pr}\, \widehat Z_m \right) \, \mathbb{Q} \,  \right|_{\mathbb{Q} =0} =0,
\nonumber
\eeq
where the operator 
$
\widehat Z_m \doteq Z_{m} (u_{n,m},u_{n,m\pm 1},u_{n\pm 1,m},...;\alpha_2,\alpha_1) \,  \partial_{u_{n,m}}
$
is obtained from $\widehat Z_n$ under $n \leftrightarrow m$, $\alpha_1 \leftrightarrow \alpha_2$, so that
$$
\left. {\rm pr}\, \widehat Z_m \, \mathbb{Q}\,  \right|_{\mathbb{Q} =0}= 
b \, g_{n,m}(u_{n,m},u_{n\pm 1,m},u_{n,m\pm 1},...;\alpha_2,\alpha_1),
$$
being $b$ a constant. So 
\beq
\widehat Z_{n,m} \doteq  \frac{1}{a}\,\widehat Z_n - 
\frac{1}{b}\, \widehat Z_m \nonumber
\eeq
is a symmetry of $\mathbb{Q} =0$.
\end{theorem}
Using this theorem it is easy to show that from the master symmetry  (\ref{nvol2}) we can construct a  generalized symmetry, given by
\bea 
\nonumber
\frac{d  u_{n,m}(\epsilon)}{d  \epsilon} &=&\frac{4 \, n\,
  (u_{n,m}-u_{n-1,m}+\alpha_0)\,
  (u_{n,m}-u_{n+1,m}-\alpha_0)}{u_{n+1,m}-u_{n-1,m}+2 \, \alpha_0}+ \\ \nonumber
&& +\, \frac{4 \, m \, (u_{n,m}-u_{n,m-1}+\beta_0)\,
  (u_{n,m}-u_{n,m+1}-\beta_0)}{u_{n,m+1}-u_{n,m-1}+2 \, \beta_0} .
\eea
The above symmetry has been implicitly used, together with point symmetries, by Nijhoff and Papageorgiou
\cite{np} to perform the similarity reduction of the lSKdV equation and get a
discrete
analogue of the Painlev\'e II equation.

\section{Conclusions} \label{S4}

In this paper we have constructed by group theoretical methods the   symmetries of the lSKdV equation. 
The Lie point symmetry algebra provides a M\"obius transformation of the
dependent variable.
 The generalized symmetries are obtained  in a constructive way by considering
 the spectral problem associated with the lSKdV equation. As was shown in
 \cite{frank1,frank} we can associate with this integrable lattice equation an
 asymmetric discrete Schr\"odinger spectral problem with two potentials. Since
 it is not easy to construct its associated evolution equations we have
 transformed the obtained spectral problem into the standard discrete
 Schr\"odinger spectral problem, whose corresponding evolution equations are the Toda and Volterra hierarchies.  
In this way we have constructed  two classes of integrable symmetries, associated with the $n$ and $m$ components of the Lax pair and we have obtained  symmetries starting from
the master symmetries. 
So we have obtained in a group theoretical framework the nonlinear
reductions
considered in \cite{np}

We left to future work  to prove if the two classes of isospectral symmetries are
independent and whether the symmetries obtained from the master symmetries are
integrable or not.
Another open problem is  the construction of the recursive
operator for the lSKdV symmetries.

\section*{Acknowledgements}
D.L.  and M.P. were partially supported by  PRIN Project ``SINTESI-2004'' of the  Italian Minister for  Education and Scientific Research. This research is part of a joint Italian Russian research project ``Classification of integrable discrete and continuous
models" financed by Consortium EINSTEIN and Russian Foundation for Basic Research.

\end{document}